# Creating Magnetic Fields in excess of 1000 T by Misoriented Stacking in a Graphene Bilayer

Wen-Yu He, Ying Su, Mudan Yang, Lin He*

Department of Physics, Beijing Normal University, Beijing, 100875, People's Republic of China

**It is well established that some kinds of lattice deformations in graphene monolayer, which change electron hopping in sublattice and affect in-plane motion of electrons, may induce out-of-plane pseudo-magnetic fields as large as 100 T. Here, we demonstrate that stacking misorientation in graphene bilayers mimics the effect of huge in-plane pseudo-magnetic fields greater than 1000 T on the interlayer hopping of electrons. As well as addressing the similarity between the effect of in-plane pseudo-magnetic fields and twisting on the electronic band structure of Bernal graphene bilayer, we point out that in-plane magnetic fields (or twisting) could modify the low-energy pseudospin texture of the graphene bilayer (the pseudospin winding number is reduced from 2 to 1), thereby changing the chiralities of quasiparticles from that of spin 1 to spin 1/2. Our results illustrate the possibility of controllably manipulating electronic properties of Bernal graphene bilayer by introducing the in-plane magnetic field or twisting.**

Graphene's novel electronic properties are a consequence of its unique crystal structure [1]. One can relatively easy to tune its electronic spectra because of the fact that graphene is a single-atom-thick membrane of carbon [2-4]. This ultimate thin film is always wrinkled to certain degree [5,6]. Recently, it has been shown that a strain-induced in-plane hopping modulation of graphene's honeycomb lattice affects the low-energy Dirac fermions like an out-of-plane pseudo-magnetic field [7,8]. Such a lattice deformation can result in partially pseudo-Landau levels at discrete energies in the band structure of graphene [9-14]. Many unusual properties and interesting behaviors are predicted to be realized in the deformed graphene [15-23]. For example, the time-reversal symmetrical pseudo-magnetic field may generate a large Zeeman-type spin splitting in graphene nanoribbons by introducing a spin-orbit coupling, thereby suggesting a non-magnetic approach for manipulating the spin degrees of freedom of electrons [22,23]. These interesting results are among the reasons that the pseudo-magnetic fields have attracted so much attention [7-23].

In this letter, we address the possibility to generate in-plane pseudo-magnetic fields in graphene system, and we show that in-plane pseudo-magnetic fields greater than 1000 T can be obtained in graphene bilayer by stacking misorientations. A twisting of Bernal graphene bilayer deflects the interlayer hopping of quasiparticles in analogy to the Lorentz force, with a small twisted angle producing a previously inaccessible huge in-plane pseudo-magnetic field. We further point out that in-plane magnetic fields (or twisting) could modify the low-energy pseudospin texture of the graphene bilayer, thereby changing the chiralities of quasiparticles and their chiral tunneling dramatically. These features seem to be promising for designing future electronic devices based on graphene bilayers.

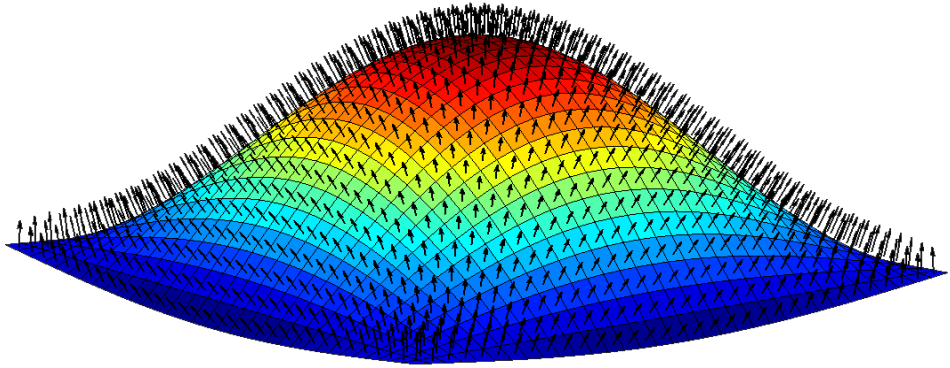

**Figure 1** (color online). The curved surface represents a strained graphene sheet. We can divide the graphene sheet into small grids and the black arrows are the normal vectors of the local grids. The strain-induced pseudo-magnetic field in graphene monolayer has opposite signs for charge carriers in the two low-energy valleys, $K$ and $K'$, and the pseudo-magnetic field in one of the valleys is always parallel to the normal vectors of the local grids.

In deformed graphene monolayer, the strain-induced pseudo-magnetic field preserves time reversal symmetry and has opposite signs for charge carriers in the two low-energy valleys, $K$ and $K'$ [7,8,22]. A gauge field in one of the two valleys can be expressed by a two-dimensional strain field $u_{ij}(x,y)$ as

$$A = \frac{\beta}{a}\begin{pmatrix} u_{xx} - u_{yy} \\ -2u_{xy} \end{pmatrix}, \qquad (1)$$

where $a$ is on the order of the C-C bond length, $2 < \beta = -\partial \ln t / \partial \ln a < 3$, and $t$ is the nearest-neighbour hopping parameter [8]. We can immediately obtain the pseudo-magnetic field $B_S$ according to

$$\vec{B}_S = \nabla \times A = -\frac{\beta}{a}\left(\frac{2\partial u_{xy}}{\partial z}, \frac{\partial(u_{xx}-u_{yy})}{\partial z}, -\frac{2\partial u_{xy}}{\partial x} - \frac{\partial(u_{xx}-u_{yy})}{\partial y}\right). \tag{2}$$

The first and second terms of Eq. (2) represent the in-plane components of the pseudo-magnetic field $B_{Sx}$ and $B_{Sy}$, respectively. To obtain a non-zero in-plane pseudo-magnetic field, the two-dimensional strain field should have a non-zero component in the out-of-plane direction. However, for any wrinkled graphene monolayer, the local strain field always has zero component along the normal vector of the local graphene sheet, as shown in Figure 1. Therefore, it is impossible to realize in-plane pseudo-magnetic fields in graphene monolayer. This result is not difficult to be understood since that the deformations of graphene's lattice only deflect the in-plane motion of quasiparticles, which mimics the effect of in-plane Lorentz force generated by a perpendicular magnetic field. An in-plane magnetic field only affects the motion of quasiparticles in the out-of-plane direction, thereby, the in-plane pseudo-magnetic field can be created only when the quasiparticles have the possibility of hopping in the perpendicular direction of graphene sheet.

The most simple structure realization of the in-plane pseudo-magnetic field is graphene bilayer where the hopping between graphene planes is included. For the purpose of this work, we consider the effect of an in-plane magnetic field on the band structure of Bernal graphene bilayer, as shown in the schematic structure of Figure 2(a). The effects of a real magnetic field on charge carriers contain two parts: the orbital field and the Zeeman field. Usually, the pseudo-magnetic field does not have the Zeeman effect of the magnetic field [22,23]. Therefore, we only consider the orbital effect of the magnetic field. Subject to

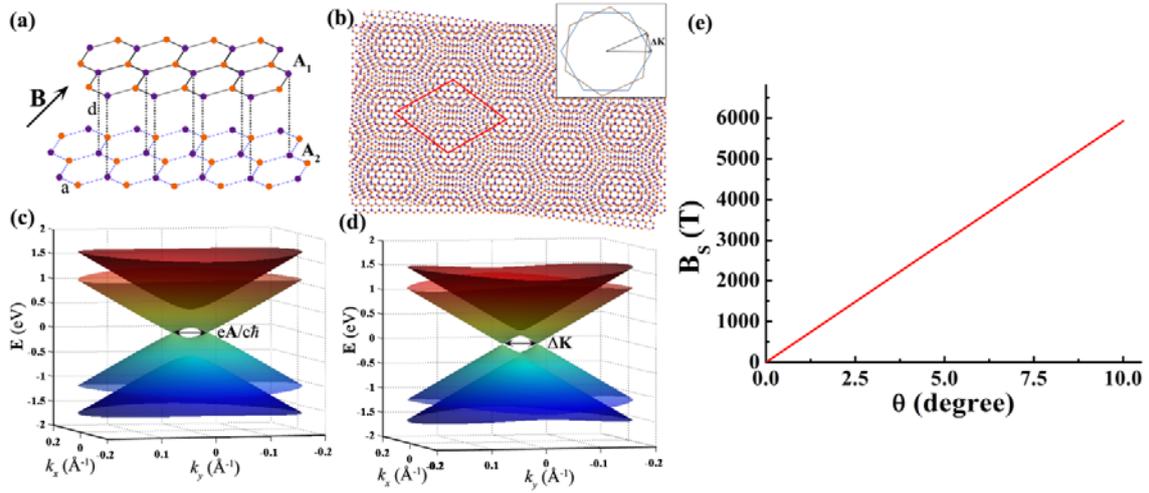

**Figure 2** (color online). (a) Schematic structure of Bernal graphene bilayer in the parallel magnetic field **B** applied along the -$x$ direction. $A_1(A_2)$ and $B_1(B_2)$ are the two sublattices of the lower (upper) lattice. d = 3.3 Å is the distance between the layers. (b) Structural model of two misoriented graphene layers with a twist angle that satisfies a condition for commensurate periodic structure leading to Moiré patterns. The inset shows schematic Dirac points of a twisted graphene bilayer. The separation of the Dirac cones Δ**K** in the two layers is attributed to the rotation between the graphene layers. (c) Electronic band structure of Bernal graphene bilayer in a parallel magnetic field **B** = 1273 T applied along the -$x$ direction. (d) Electronic band structure of twisted graphene bilayer with a twisted angle θ = 2.15°. Both the in-plane magnetic field and the twisting split the parabolic spectrum of Bernal graphene bilayer into two low-energy Dirac cones. (e) The magnitude of the in-plane pseudo-magnetic field induced by twisting as a function of the twisted angle. A small twisted angle in graphene bilayer generates a previously inaccessible huge in-plane pseudo-magnetic field.

in-plane orbital fields, an electron tunneling between the layers will obtain an excess in-plane momentum $\Delta p_y = eA/c = \hat{y} eBd/c$ because of the Lorentz force $\mathbf{F} = -\frac{e}{c}[\mathbf{v} \times \mathbf{B}]$ (here the parallel magnetic field $\mathbf{B}$ is applied along the -$x$ direction and we choose the gauge $\mathbf{A} = \hat{y}Bz$, $d$ is the distance between the layers, $e$ is the electron charge, and $c$ is the velocity of light) [24]. Applying the Peierls substitution $p \rightarrow p+e\mathbf{A}/c$, then we obtain the Hamiltonian of Bernal graphene bilayer in a parallel magnetic field [24,25]

$$H(\vec{k}) = \begin{pmatrix} \frac{v_F}{\hbar}\left[\left(\hbar\vec{k} - \frac{1}{2}\Delta\vec{p}\right)\cdot\vec{\sigma}\right] & t_1 I_A \\ t_1 I_A & \frac{v_F}{\hbar}\left[\left(\hbar\vec{k} + \frac{1}{2}\Delta\vec{p}\right)\cdot\vec{\sigma}^*\right] \end{pmatrix}. \quad (3)$$

Here $v_F = \frac{3}{2}at$ is the Fermi velocity, $\vec{\sigma} = (\sigma_x, \sigma_y)$ are the pauli matrices, and the term $t_1 I_A$ represents the interlayer coupling with the matrix $I_A = \frac{1}{2}(I + \sigma_z)$ connecting sublattices A of the adjacent graphene layers. Since the in-plane magnetic field differs the vector potential of each layer, it splits each parabolic band touchings in Bernal graphene bilayer into two Dirac cones, i.e., the Dirac point is shifted to $-\frac{1}{2}\Delta p_y$ of layer 1 and $\frac{1}{2}\Delta p_y$ of layer 2, as shown in Figure 2(c). Consequently, two intersections of the saddle points along the two Dirac cones appear in the low-energy band spectrum. The saddle points will result in two van Hove singularities in the density of states.

A stacking misorientation mimics the effect of in-plane magnetic fields on the band structure of Bernal graphene bilayer. For a graphene bilayer with a twisting, the Dirac points of the two layers no longer coincide, as shown in Figure 2(b). The relative shift of the Dirac points on the different layers in the momentum space is |Δ**K**| = 2|**K**|sin($\theta$/2), where $\theta$ is the twisted angle and $\mathbf{K} = \left(\frac{2\pi}{3a}, \frac{2\pi}{3\sqrt{3}a}\right)$ is the reciprocal-lattice vector [26-32].

The Hamiltonian of the twisted graphene bilayer can be written as

$$H(\vec{k}) = \begin{pmatrix} v_F\left(\left(\vec{k} - \frac{1}{2}\Delta\vec{K}\right) \cdot \vec{\sigma}\right) & t_1 I_A \\ t_1 I_A & v_F\left(\left(\vec{k} + \frac{1}{2}\Delta\vec{K}\right) \cdot \vec{\sigma}^*\right) \end{pmatrix}. \qquad (4)$$

Figure 2(d) shows electronic spectrum of a twisted graphene bilayer in the proximity of one of the two valleys. For convenience, we choose appropriate coordinate to make $\Delta K_x = 0$ and $\Delta K_y = |\Delta\mathbf{K}|$. Such a band structure of the twisted graphene bilayer was confirmed experimentally by scanning tunnelling spectroscopy [14,28-30,33,34], Raman spectroscopy [35,36], and angle-resolved photoemission spectroscopy [37]. The similarity of the Hamiltonian (3) and (4) of the two systems (and the similarity between the low-energy band spectra shown in Figure 2(c) and 2(d)) ensures that it is reasonable to treat a slight twisting in graphene bilayer as an in-plane pseudo-magnetic field. The in-plane pseudo-magnetic field induced by the twisting can be estimated by $B_S = \frac{2\hbar c K \sin(\theta/2)}{de}$ according to the Hamiltonian (3) and (4). As shown in Figure 2(e), a small twisted angle produces a previously inaccessible huge in-plane pseudo-magnetic field because of the fact that the effect of a twisting on the interlayer hopping of quasiparticles is much stronger than that of a Lorentz force generated by a real magnetic field. Additionally, the scheme of producing such a great pseudo Lorentz force from nonuniform interlayer hopping should

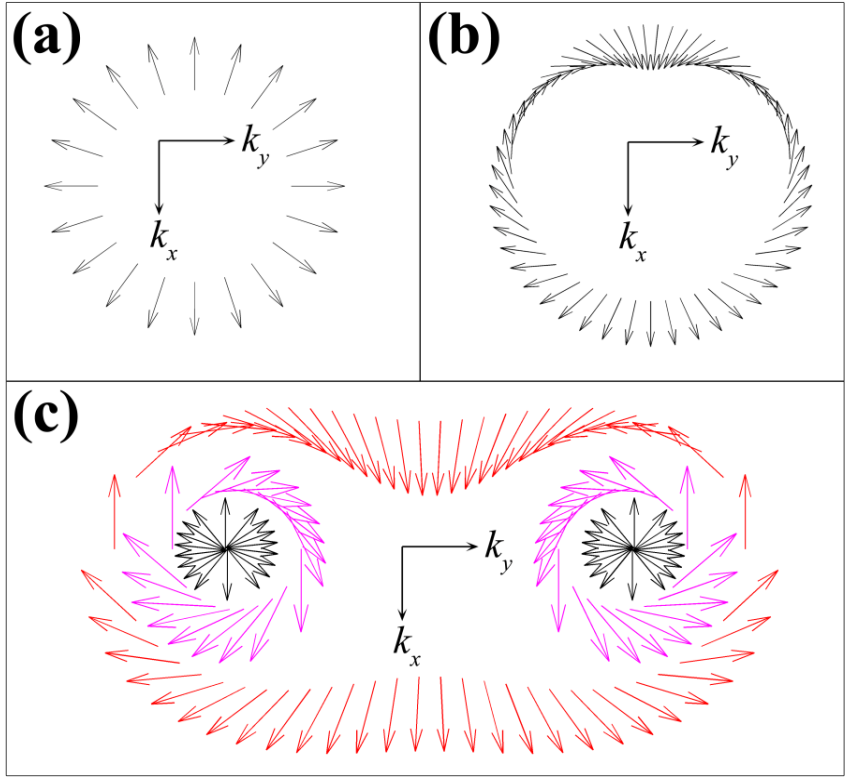

**Figure 3** (color online). (a) A typical two-dimensional pseudospin distribution for electronic eigenstates in graphene monolayer on an equienergy contour of the upper Dirac cone. (b) Pseudospin distribution for eigenstates in Bernal graphene bilayer on an equienergy contour above the charge neutrality point. (c) Pseudospin distribution for eigenstates in Bernal graphene bilayer in a parallel magnetic field of 1273 T on different equienergy contours. The black color denotes the pseudospin vectors at $E = 0.27$ meV; purple denotes the pseudospin vectors at $E = 16.4$ meV; red denotes the pseudospin vectors at $E = 38.3$ meV. Interestingly, the pseudospin distribution forms a pattern somewhat resembling an owl's eyes.

also be feasible for graphene multilayers with stacking misorientation.

Both the in-plane magnetic field and the twisting change the dispersion for low-energy quasiparticles in Bernal graphene bilayer from quadratic to linear. We will demonstrate subsequently that such a variation of the electronic spectrum is nontrivial for the electronic properties of the graphene system. Figure 3 shows the two-dimensional pseudospin texture of graphene monolayer, Bernal graphene bilayer, and Bernal graphene bilayer in a parallel magnetic field (or twisted graphene bilayer). The pseudospin vector of graphene system is defined from the relative phase of the two components of their wave function (a two-component spinor) [29,41-43]. For graphene monolayer, the two components are composed of the Bloch sums of orbitals localized on the two sublattices $A_1$ and $B_1$. For graphene bilayer, including Bernal graphene bilayer and twisted graphene bilayer, the two components represent the Bloch sums of localized orbitals on each of the two sublattices $A_1$ and $B_2$ (here $A_1$, $B_1$ and $A_2$, $B_2$ are denoted as the two sublattices in layer 1 and 2, respectively) [29,41-43]. The pseudospin winding number $n_w$, which is defined as the number of times the pseudospin vector rotates when the electronic wave vector undergoes one full rotation around the charge neutrality point [41], of Bernal graphene bilayer is 2. The in-plane magnetic field (or twisting) changes its low-energy pseudospin distribution dramatically. In each of the Dirac cone, as shown in Figure 3(c), $n_w$ becomes 1, which is identical to that of graphene monolayer. The number of $n_w$ determines the behavior of quantum Hall effect in graphene systems [41]. It changes from 1 to 2 when the Fermi energy cross the van Hove singularity, at which the two Dirac cones merge into a single parabolic-like band spectrum. Therefore, it is expected to observe a crossover of the quantum Hall effect at the van Hove singularities in such a unique system. Below the van Hove singularity, the quantum Hall effect of the system should be similar to the monolayer case except for an extra twofold degeneracy due to the Dirac point splitting. Above the van Hove singularity, the quantum Hall effect of the system should behave as that of Bernal graphene bilayer. Our analysis based on the pseudospin winding number of the system consists well with the numerical result reported in Refs.44 and 45.

In addition to the variation of the pseudospin winding number, the in-plane magnetic field (or twisting) also changes the chiralities of quasiparticles in Bernal graphene bilayer substantially. In Bernal graphene bilayer under zero magnetic field, quasiparticles exhibit chiralities that resemble those associated with spin 1 [43]. However, the in-plane magnetic field (or twisting) can change the chiralities of quasiparticles around the charge neutrality points to those associated with spin 1/2. Therefore, the transmission probability for normally incident chiral fermions around the charge neutrality points is expected to be tuned from perfect reflection to perfect tunneling, or vice versa [29,43]. To further illustrate the effect of in-plane magnetic fields on the chiralities of quasiparticles around the charge neutrality points, we calculate the chiral tunneling in a Bernal graphene bilayer in different magnetic fields. We consider the chiral fermions with incident energy $E$ penetrating through a potential barrier with height $E + \Delta U$ and width $D$ (see Supplemental Material [46] for details). The potential barrier is infinite along the $y$ axis and it is further assumed to have a rectangular shape to reduce scattering of quasiparticles between the two valleys in graphene [28,43]. The electron wave function of Bernal graphene bilayer is the vector $\Psi = \left( \psi_1^A, \psi_1^B, \psi_2^A, \psi_2^B \right)$, where the subscript 1 and 2 enumerate the layers [24,47]. By written

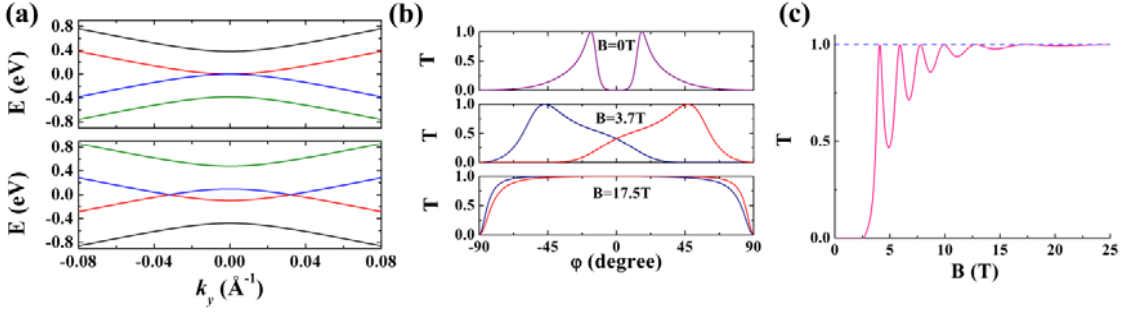

**Figure 4** (color online). (a) Upper panel: a section view of band structure of Bernal graphene bilayer along $k_y$ for $k_x = 0$. Lower panel: a section view along the two Dirac points of band structure of Bernal graphene bilayer in a parallel magnetic field of 1273 T. (b) Transmission probability of quasiparticles in Bernal graphene bilayer as a function of the incident angle. The quantum tunneling depends very sensitively on the magnitude of the parallel magnetic fields. (c) The transmission probability as a function of the in-plane magnetic field for normally incident quansiparticles. The remaining parameters in the calculation are the incident energy $E = 0.001$ meV, $\Delta U = 0.01$ meV, and the width of barrier $D = 10$ μm.

out the wave functions in the three different regions of the tunneling system (i.e., the left of the barrier, inside the barrier, and the right of the barrier), then it is straightforward to solve this tunneling problem (see Supplemental Material [46] for details).

Figure 4 shows the effect of in-plane magnetic field on the band structure and chiral tunneling of Bernal graphene bilayer. The angle dependent transmission probability, as shown in Figure 4(b), depends sensitively on the in-plane magnetic field. The most striking result of the tunneling problem is the variation of transmission probability at $\varphi = 0$, i.e., for normally incident fermions. In zero magnetic field, the massive chiral fermions are always perfectly reflected for a sufficiently wide barrier for normal incidence. A parallel magnetic field enhances the transmission probability at $\varphi = 0$ and even can change the system from perfect reflection to perfect tunneling for normally incident fermions, as shown in Figure 4(b) and 4(c). This result directly demonstrated that the in-plane magnetic fields (or twisting) change the chiralities of quasiparticles around the charge neutrality points from those associated with spin 1 to those associated with spin 1/2. With considering that ultrahigh magnetic fields up to 160 T could be obtained experimentally [48], the predicted effect in this paper is expected to be realized by transport measurements [49,50] in the near future.

In conclusion, we demonstrate that a twisting of Bernal graphene bilayer produces huge in-plane pseudo-magnetic fields greater than 1000 T, which could provide a platform to study charge carriers in previously inaccessible high magnetic field regimes. We further point out that the pseudospin winding number and chiralities of quasiparticles of the graphene bilayer can be tuned by in-plane magnetic fields (or twisting). These results indicate that the graphene bilayer is a unique system with tunable internal degrees of freedom. The ability to tune these internal degrees of freedom is promising for designing future electronic devices.

## Acknowledgements

We are grateful to National Science Foundation (Grant No. 11374035, No.

11004010) ,National Key Basic Research Program of China (Grant No. 2013CBA01603, No. 2014CB920903), and the Fundamental Research Funds for the Central Universities.

*Email: helin@bnu.edu.cn.